\documentclass{article}

\usepackage{arxiv}

\usepackage[breaklinks=true]{hyperref}       
\usepackage{booktabs}       
\usepackage{amsfonts}       
\usepackage{amsmath}
\usepackage{nicefrac}       
\usepackage{graphicx}
\usepackage{natbib}
\usepackage{doi}
\usepackage{subcaption}

\title{Wall-resolved large eddy simulations of a pitching airfoil incurring in deep dynamic stall}

\date{} 					

\author{
	\href{https://orcid.org/0000-0002-6664-4170}{\includegraphics[scale=0.06]{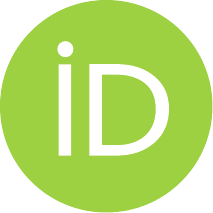}\hspace{1mm}Giacomo Baldan} \\
	Department of Aerospace Science and Technology \\
	Politecnico di Milano\\
	Milano, 20156 \\
	\texttt{giacomo.baldan@polimi.it} \\
	\And
	\href{https://orcid.org/0000-0001-6432-2461}{\includegraphics[scale=0.06]{orcid.pdf}\hspace{1mm}Alberto Guardone} \\
	Department of Aerospace Science and Technology \\
	Politecnico di Milano\\
	Milano, 20156 \\
	\texttt{alberto.guardone@polimi.it} \\
}

\renewcommand{\headeright}{}
\renewcommand{\undertitle}{}
\renewcommand{\shorttitle}{A deep neural network physics-based reduced order model for dynamic stall}

\hypersetup{
pdftitle={A deep neural network physics-based reduced order model for dynamic stall},
pdfauthor={Giacomo Baldan, Alberto Guardone},
}

\begin{document}
\maketitle

\begin{abstract}
	This study investigates the flow evolution around a sinusoidal pitching NACA 0012 airfoil, defined by the National Advisory Committee for Aeronautics (NACA),  undergoing deep dynamic stall using a wall-resolved large eddy simulation (LES) approach.
	Numerical results are assessed against experimental data from Lee and Gerontakos (2004) at Reynolds number Re = 135\,000 and reduced frequency $k$ = 0.1.
	A comprehensive analysis of the computational model span size is presented, highlighting the  requirement for a span-to-chord ratio of at least one to correctly capture the dynamic stall vortex physics in the downstroke phase.
	Furthermore, a comparative assessment with state-of-the-art Reynolds-Averaged Navier-Stokes (RANS), hybrid RANS/LES, and the experimental data is carried out.
	All the numerical models concur to the same flow behavior and exhibit similar differences with the experiments.
\end{abstract}


Dynamic stall is a transient unsteady phenomenon impacting airfoils subjected to rapid variations in the angle of attack.
It deviates significantly from the well-established theory of static stall, where an abrupt reduction in lift occurs once a limit value of the angle of attack, the stall angle, is exceeded.
In dynamic stall, the separation of the attached boundary layer triggers the formation of the dynamic stall vortex (DSV) that convects downstream along the suction surface of the airfoil.
The subsequent development and interaction of the DSV with the boundary layer leads to complex flow separation phenomena~\citep{Smith2020}. 
Dynamic stall induces flow separation over a substantial portion of the airfoil at angles of attack exceeding the critical static stall limit.
The extensive flow separation results in a drastic, and often delayed, reduction in lift, combined with a significant increase in drag and pitch down moment.
Dynamic stall can have a profound impact on the performance and stability of aerodynamic structures.
Notably, dynamic stall plays a crucial role in several key areas: helicopter rotor blades, wind turbines, and high-performance aircraft maneuvers.
The occurrence of dynamic stall limits the maximum speed of conventional rotorcraft.

\begin{figure*}
	\centering
	\includegraphics[width=0.6\textwidth]{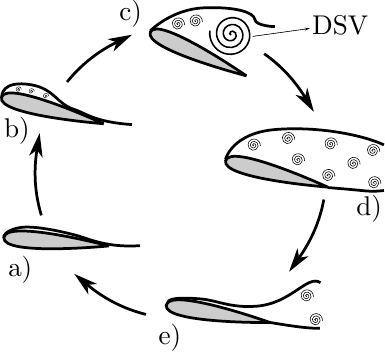}
	\caption{Sketch of dynamic stall over a pitching airfoil: a) attached flow at small angles of incidence; b) laminar separation bubble generates increasing the incidence of the profile; c) dynamic stall vortex detaches and convects backward; d) complete stalled flow over the airfoil during the downstroke phase; e) flow reattachment at reduced angle of attack.}
	\label{img_dynamic_stall_sketch}
\end{figure*}

With reference to Figure~\ref{img_dynamic_stall_sketch}, it is possible to identify five main phases of the dynamic stall. 
a) During the initial portion of the cycle, the airfoil experiences attached flow due to the low angles of incidence.
b) As the angle of attack increases throughout the upstroke phase, a laminar separation bubble forms on the airfoil. Concurrently, the boundary layer transitions from laminar to turbulent flow, initiating at the trailing edge and progressing downstream.
c) The interaction between the boundary layer and the laminar separation bubble triggers the formation of the dynamic stall vortex, which eventually convects rearward.
d) Upon reaching the maximum angle of incidence and at the beginning of the downstroke phase, the flow over the airfoil suction side becomes fully stalled.
e) Reattachment starts at a reduced angle of attack due to the negative pitch rate, which increases the effective angle of attack.

The complex and transient nature of dynamic stall presents a significant obstacle in modeling, simulation, and design of aerodynamic bodies operating in dynamic stall conditions~\citep{Gardner2023, Baldan2024c}.
Research efforts primarily focus on developing robust high-fidelity Computational Fluid Dynamics (CFD) tools, and low-fidelity reduced order models (ROMs) for implementing flow control strategies.
Indeed, a deeper understanding of dynamic stall and its impact allows for improving the design and control of aerodynamic bodies~\citep{Gardner2019, Hariharan2014}. 
This will ultimately lead to \textit{e.g.}~safer and more efficient flight, improved wind energy generation, and robust performance across diverse operating conditions.

A large number of contributions in the open literature rely on the Unsteady Reynolds-Averaged Navier-Stokes (URANS) models and hybrid RANS/LES techniques to study dynamic stall~\citep{Kim2016, Karbasian2016, Khalifa2021, Khalifa2023, Geng2018, Zanotti2013, Zanotti2014a, Zanotti2014b, Avanzi2022, Avanzi2021, Gardner2020, Baldan2024a, Baldan2024b}.
Most of these works conclude that the discrepancies between experimental data and numerical simulations rely on the adopted turbulence model to close the RANS system.
A vast number of turbulence models have been adopted ranging from one equation Spalart-Allmaras formulation~\citep{Spalart1992} to full Reynolds stress model (RSM)~\citep{Launder1975}, and including the well-known $k-\epsilon$~\citep{Launder1974} and $k-\omega$ SST~\citep{Menter1994}.
Also, transition models are included, for instance, the recent $k-\omega$ SST with intermittency equation~\citep{Menter2004}.
Regarding hybrid strategies, common choices are detached eddy simulations (DES)~\citep{Spalart1997} and their improved versions (DDES, IDDES).
More recent approaches are scale adaptive simulations (SAS)~\citep{Menter2010} and stress-blended eddy simulations (SBES)~\citep{Menter2018}.

Large Eddy Simulations (LES) have been applied to investigate dynamic stall.
In particular, several aspects have been numerically investigated for finite wings ranging from aspect ratio~\citep{Hammer2022} to sweep angle~\citep{Hammer2023} and compressibility effects~\citep{Benton2020}.
Diverse studies concern nominal infinite wings obtained by imposing periodic boundary conditions in the spanwise direction.
According to the best practice DES, presented in~\citet{Shur1999}, a spanwise size of at least one chord is required to reproduce the DSV accurately.
Indeed, it is well known from experimental investigations that the characteristic dimension of the DSV is on the order of the airfoil chord length when incurring in deep dynamic stall~\citep{McCroskey1981}.
Despite the considerations mentioned, a significant portion of LES literature employs extruded airfoil geometries with spanwise lengths ranging from only 5\% to 20\% of the chord size.
In~\citet{Guillaud2018}, a pitching NACA0012 airfoil at Re~$=2\cdot10^4$ is investigated using a numerical model with spanwise length of $0.5c$.
In particular, the analysis compared the reduced frequency of the aerodynamic coefficients drop delay to the static case.
Studies on dynamic stall phenomenology and control have been proposed by~\citet{Benton2019a, Benton2019b, Visbal2018b} and leverage high-frequency control to mitigate the intense burst of the aerodynamic loads due to the Leading Edge Vortex (LEV) detachment.
These investigations also focused on the effect of the Reynolds number, which is on the order of $10^6$, on the laminar separation bubble (LSB).
However, the wing spanwise size in all contributions is limited to $0.05c$, $0.1c$ and $0.2c$.
Recently, \citet{Lee2024} proposed to replicate LES results presented in the literature by reproducing a ramp-up motion over a straight wing with spanwise size equal to $0.1c$ and a sinusoidal pitching test case using a different geometry with span-to-chord ratio of 0.02.
The same authors analyzed, in~\citet{Lee2024b}, the effects of Reynolds and Mach numbers on dynamic stall onset for harmonically pitching airfoils with LES but still limiting the span size of the model to $0.1c$.
\citet{Lee2024} work agrees with the reference data of~\citet{Gupta2019}, while \citet{Lee2024b} results show significant discrepancies in the downstroke phase compared to~\citet{Benton2019a}.
\citet{Visbal2018a}, at the beginning of the work, focused on the span effect starting from $0.1c$ and up to $1.6c$ in a ramp-up motion.
They analyzed the discrepancies in the loads and the span-averaged pressure distribution over the wing. 
They concluded that there is a dependency on the span size after the DSV interaction with the Trailing Edge Vortex (TEV).
Despite the initial considerations on the model span extension, the core investigation focused on the LSB and DSV generation, and all the following displayed results refer to the $0.1c$ case.
Another contribution by the same authors, \citet{Visbal2019a}, concerns the numerical simulations of typical experimental setups.
Namely, the aspect ratio of the wing section, together with end walls and end tips, is considered.
They provide a quantification of the discrepancy magnitude among all the configurations, but they do not report any comparison with actual experimental data.
Less recent contributions tried to match experiments with LES.
\citet{Patil2018} delves into analyzing an H-Darrieus rotor to match wind tunnel data over an entire revolution.
They captured the attached flow behavior but not the stalled region.
Their analysis limited the span size to $0.05c$.
The contributions of \citet{Badoe2019, Kasibhotla2014} investigate pitching NACA0012 airfoils at $2\cdot 10^4$ and $10^5$ Reynolds number using geometries with spanwise lengths of $0.25c$ and $0.18c$, respectively.

Despite recent efforts, resorting to LES or DES models do not reduce the discrepancy observed between numerical simulations and experimental data.

The present work presents a numerical investigation of a NACA0012 airfoil undergoing a full sinusoidal pitching motion at a Reynolds number of 135\,000 and a reduced frequency of $k = 0.1$, using a wall-resolved large eddy simulation approach. 
The goal is to analyze the influence of spanwise length on the aerodynamic behavior during the upstroke and downstroke phases during a full sinusoidal motion.
Span sizes ranging from $0.2c$ to $1.2c$ are examined to assess the dependence of the dynamic stall evolution on the span size.
Numerical results are compared with the experimental data proposed by~\citet{Lee2004}.
A comparison is presented with current state-of-the-art RANS and hybrid RANS/LES models.
This comparative analysis aims to elucidate whether the observed discrepancies between the numerical results and the experimental data stem from the turbulence model employed or from other factors, such as model roughness.

Section~\ref{sec_numerical_setup} details the computational grid and numerical methods utilized.
Section~\ref{sec_results} presents the results, emphasizing the impact of the spanwise extension on characteristic dynamic stall phenomena.
Finally, Section~\ref{sec_conclusion} reports the conclusions drawn from the study.

\section{Numerical setup}\label{sec_numerical_setup}
	

\begin{figure*}[tb]
	\centering
	\includegraphics[width=\textwidth]{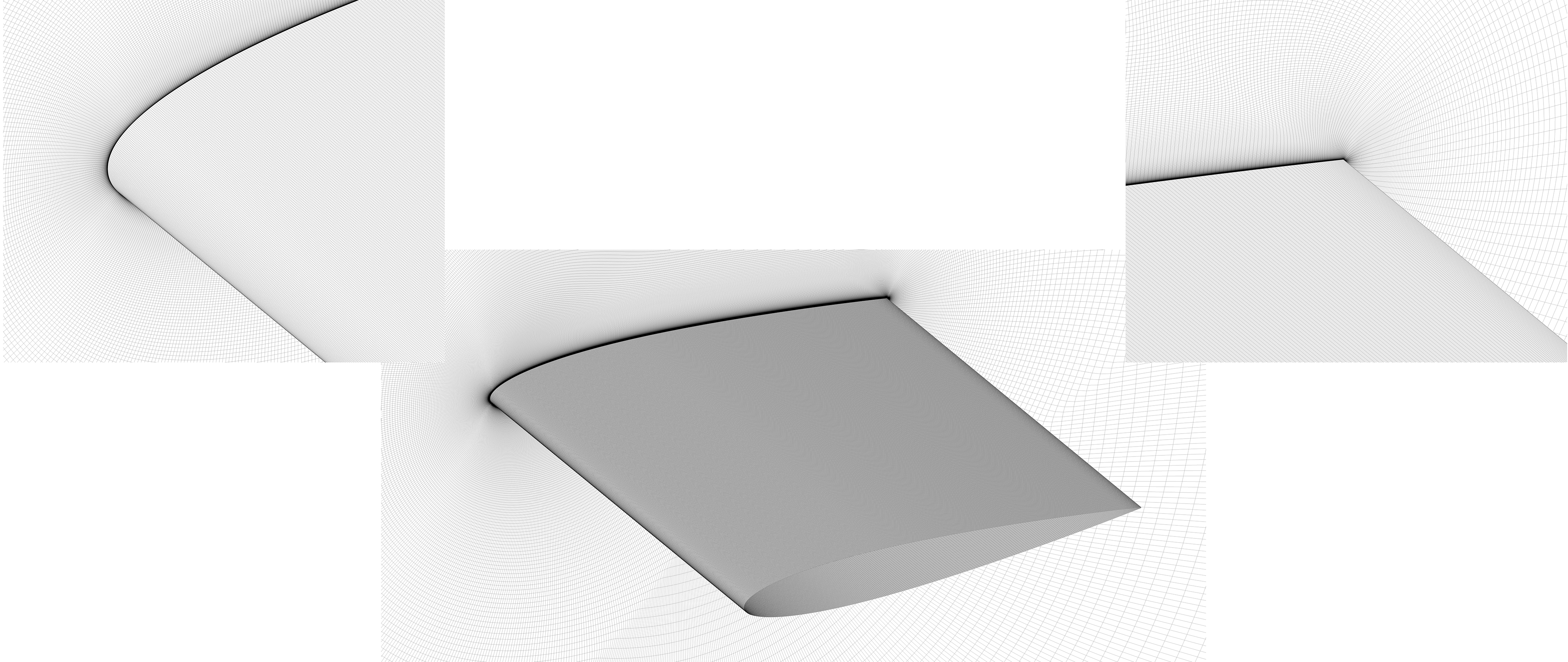}
	
	\caption{Details of the three-dimensional mesh of the wing section with span-to chord ratio of 1.2.}
	\label{img_mesh_3d}
\end{figure*}

The present numerical simulations replicate the experimental conditions presented in~\citet{Lee2004}.
A NACA0012 airfoil underlying a sinusoidal pitching motion defined as:
\begin{align}
	\begin{split}
		\alpha(t) &= \alpha_0 + \alpha_s \sin(\omega t)\\
		&= 10^\circ + 15^\circ \sin \left(18.67 \, [\text{Hz}] \ t\right)
	\end{split}
\end{align}
is investigated, where $\omega$ is the pitching frequency, $\alpha_0$ is the mean angle of attach, and $\alpha_s$ is the angular oscillation amplitude.
The present simulations are carried out at Reynolds number Re~=~$1.35\cdot10^5$ and reduced frequency $k=\omega \, c / 2 V_\infty=0.1$.
The free-stream velocity is $V_\infty=14 \, \text{m/s}$ while the reference pressure is set to $P_\infty=1 \, \text{atm}$.

\subsection{Grid generation}

\begin{table*}
	\centering
	\caption{2D O-grid specifications. The size of the elements is made dimensionless by the chord size $c$.}
	\label{table_2dgrid}
	\begin{tabular}{*{8}{c}}
		\hline
		$\mathbf{N_\xi}$  & 
		$\mathbf{N_\eta}$ &  \textbf{Nodes} & 
		\textbf{Quads} & $\mathbf{\Delta \xi_\text{LE}}$ & $\mathbf{\Delta \xi_\text{TE}}$ & $\mathbf{\overline{\Delta \xi}_\text{suction}}$ & $\mathbf{\overline{\Delta \xi}_\text{pressure}}$ \\
		\hline
		934 & 334 & 311\,956 & 311\,622 & $6.67\cdot10^{-4}$ & $4.35\cdot10^{-4}$ & $1.88\cdot10^{-3}$ & $2.69\cdot10^{-3}$ \\
		\hline
	\end{tabular}
\end{table*}

\begin{table*}
	\centering
	\caption{3D grid specifications. Size of elements is reported as a function of the chord size $c$.}
	\label{table_3dgrid}
	\begin{tabular}{*{10}{c}}
		\hline
		\textbf{Span size} & $\mathbf{N_z}$ & \textbf{Nodes} & \textbf{Hexahedra} & $\mathbf{\Delta z}$\\
		\hline
		$0.2c$ &  201  & \ \;62.7M & \ \;62.1M & $1.00\cdot10^{-3}$ \\
		$0.4c$ &  401  &    125.1M &    124.3M & $1.00\cdot10^{-3}$ \\
		$0.8c$ &  801  &    249.9M &    248.6M & $1.00\cdot10^{-3}$ \\
		$1.2c$ & 1201  &    374.7M &    372.8M & $1.00\cdot10^{-3}$ \\
		\hline
	\end{tabular}
\end{table*}

The CFD mesh is generated using \textit{Fidelity Pointwise v2023.2}~\cite{Pointwise}.
To avoid the non-physical flow patterns caused by the typical C-grid strong anisotropic elements, especially near the trailing edge~\cite{Baldan2024b}, an O-grid is adopted in this research.
Table~\ref{table_2dgrid} reports the details of the two-dimensional grid, where $\xi$ is the curvilinear coordinate along the profile and $\eta$ is the direction normal to the  airfoil surface.

The O-grid is obtained through a hyperbolic extrusion in the normal direction starting from the profile surface.
The wall-normal spacing follows a geometric progression with a growth rate of 1.03.
The standard values required for Wall-Resolved Large Eddy Simulation (WRLES), typically demanding resolution values~\cite{Piomelli2002, Georgiadis2010} of $\xi^+< 40$, $\eta^+<5$, and $z^+<20$, are satisfied at the wall for all grids, even for the most demanding flow condition when the acceleration, due to the high angle of attack, increases the velocity at more than twice the free-stream value.
Thus, the first layer is positioned at $10^{-4} \ c$ over the surface. 
Another critical aspect of the mesh to capture dynamic stall phenomena is the element size at the leading and the trailing edges.
The leading edge spacing is extremely important since it covers the region where the laminar separation bubble resides during the pitch-up phase, and subsequently, the dynamic stall vortex is generated.
The same consideration can also be applied to the trailing edge since the stall also starts from the back of the profile.
Also, for this reason, the blunt trailing edge is joined with a circle that avoids the imposition of the separation point and, at the same time, increases the orthogonality of the grid, allowing faster convergence.
The element size at the nose of the airfoil is $\Delta \xi_\text{leading edge} = 6.67\cdot10^{-4}\ c$ while at the trailing edge is $\Delta \xi_\text{trailing edge} = 4.35\cdot10^{-4}\ c$.
To reduce the number of points of the meshes, the pressure side of the airfoil, which is not directly involved in the DSV advection, is discretized with a lower resolution.
Specifically, the suction side has 550 nodes, and the pressure side is limited to 385.
This results in an average spacing on the upper surface of $\overline{\Delta \xi}_\text{suction} = 1.88\cdot10^{-3}\ c$ and on the lower surface of $\overline{\Delta \xi}_\text{pressure} = 2.69\cdot10^{-3}\ c$.
In all the meshes, the farfield is placed at $50c$ from the airfoil to reduce the boundary influences.

The three-dimensional grid is obtained through a uniform extrusion of the two-dimensional O-grid in the $z$ direction.
The number of points is retrieved respecting the LES requirements~\citep{Choi2012}.
For the smallest model, extending for $0.2c$ in span, the number of points equals 201.
When moving to larger extrusions, the element size is kept the same, resulting in 401, 801, and 1201 nodes for meshes with span sizes of $0.4c$, $0.8c$, and $1.2c$, respectively.
Table~\ref{table_3dgrid} reports the details of the three-dimensional meshes.
Figure~\ref{img_mesh_3d} illustrates the three-dimensional grid with $1.2c$ span size.

\subsection{Numerical methods}

The steady incompressible RANS equations and the LES ones are solved using \textit{ANSYS Fluent 2023R2}~\citep{ANSYS}.
A second-order upwind discretization in space is adopted.
Gradients are retrieved through a least square cell-based method, and fluxes are obtained with the Rhie-Chow momentum-based formulation.
The SIMPLE method is used to solve the pressure-velocity equations in the steady framework. The SIMPLEC method is used to compute the LES solution.
The time evolution is obtained through a bounded second-order implicit time integration scheme.

Moving on to boundary conditions, the farfield tag present in the original mesh is split into two parts during the simulation evolution.
This is necessary since the mesh rigidly rotates and it is not possible to keep the same boundary condition definition.
For this reason, a marker is set for cells with positive $x$, corresponding to the coordinate around which the grid rotates and being $x$, the direction of the free-stream flow.
A velocity inlet condition is imposed to farfield faces associated to the mark, $x<0$, while a pressure outlet boundary condition is set for the others, $x \geq 0$.

At the farfield, inflow velocity is imposed together with the turbulence intensity of 0.08\% as in the wind tunnel.
A non-slip and non penetrating condition is imposed on the airfoil. 
The surfaces at the tips of the extruded O-grid features a translational conformal periodic condition to grant a nominally infinite wing.

RANS system is closed using the $k-\omega$ SST turbulence model with the addition of the $\gamma$ intermittency model~\citep{Menter2004}, in which a differential transport equation is solved for the intermittency, to better describe the transition of the boundary layer from laminar to turbulent.

Moving to large eddy simulations, WALE subgrid scale model is used since it grants the right eddy viscosity trend in near-wall areas without requiring any dumping function or transition between the bulk flow and the boundary layer.
The model provides the $\mu_{SGS}/\mu \sim \mathcal{O}(y^+)^3$ asymptotic behavior at the wall locations \citep{Chapman1986}.
In order to grant a CFL close to one, the pitching cycle is discretized into 144\,000 time steps that correspond to $2.337\cdot10^{-6} \ s$ or, adimensionalizing it, $2.181\cdot10^{-4}$ convection times.

The airfoil pitching motion is prescribed through the rigid motion of the entire grid.
The angular velocity is imposed using a user-defined expression equal to $\dot \alpha(t) = \omega \ \alpha_s \cos(\omega t)$.
The mean angle of attack $\alpha_0$ is imposed through a rigid rotation of the mesh before starting the simulation.
In order to speed up temporal convergence, the simulation is performed in three different stages. 
Firstly, a steady RANS simulation with the mesh rotated at $\alpha_0$ is performed until force coefficient convergence is reached. 
Then, the time-dependent pitching simulation is evolved.
The mesh rotation is imposed to obtain a negative pitch rate after the steady simulation~\citep{Baldan2023}.
The first pitch down quarter of the cycle is neglected.
Finally, an entire pitching cycle is computed, and data are recorded.

Following each simulation, the pressure coefficient $C_p$ and skin friction coefficient $C_f$ distributions over the wing section are extracted every 40 time steps and are mapped onto the structured mesh of the wing section using a k-d tree algorithm, which efficiently identifies the nearest neighbors.
For post-processing, the skin friction value at each point is converted to its signed magnitude.
The sign is determined based on the direction calculated from the $x$ and $y$ components of $C_f$.
The positive geometric direction is aligned with the curvilinear coordinate that extends from the leading edge to the trailing edge.
If the geometric direction and the numerically computed $C_f$ direction coincide, the sign is positive. 
Conversely, if they oppose each other, the sign is negative. 
This process allows for the identification of regions with flow separation.

\section{Results}\label{sec_results}

The present section discusses the results obtained with the numerical models described previously. 
Specifically, the effect of the span-to-chord ratio is investigated in Section~\ref{ssec_span}, while Section~\ref{ssec_other_simulations} reports the comparison of the proposed LES with other numerical approaches.

\subsection{Span-to-chord ratio dependence}\label{ssec_span}

\begin{figure*}
	\centering
	\subfloat{\includegraphics[width=0.48\textwidth]{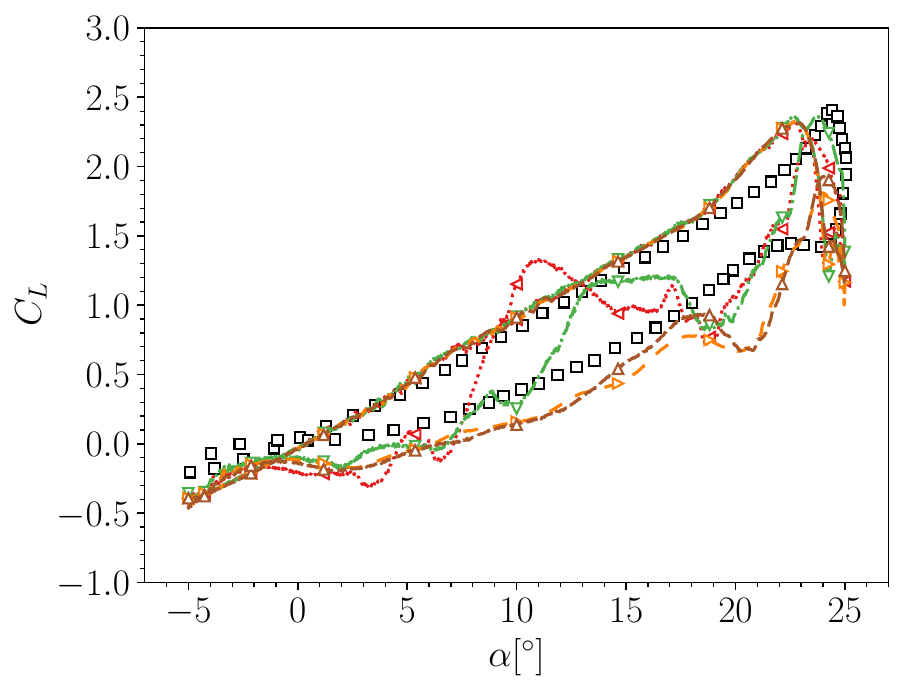}}
	\hfill
	\subfloat{\includegraphics[width=0.48\textwidth]{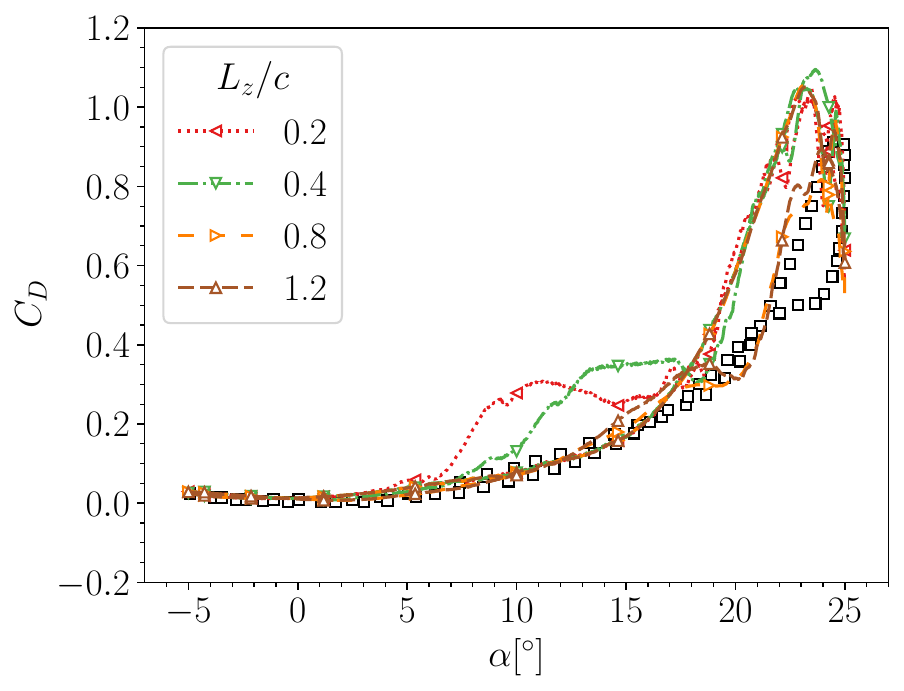}}\\
	
	\caption{Spanwise size comparison of lift and drag coefficients for the simulated cycle. Experimental data from~\citet{Lee2004}.}
	\label{img_les_span}
\end{figure*}

\begin{figure*}
	\centering
	\begin{subfigure}{\linewidth}
		\includegraphics[width=.95\linewidth]{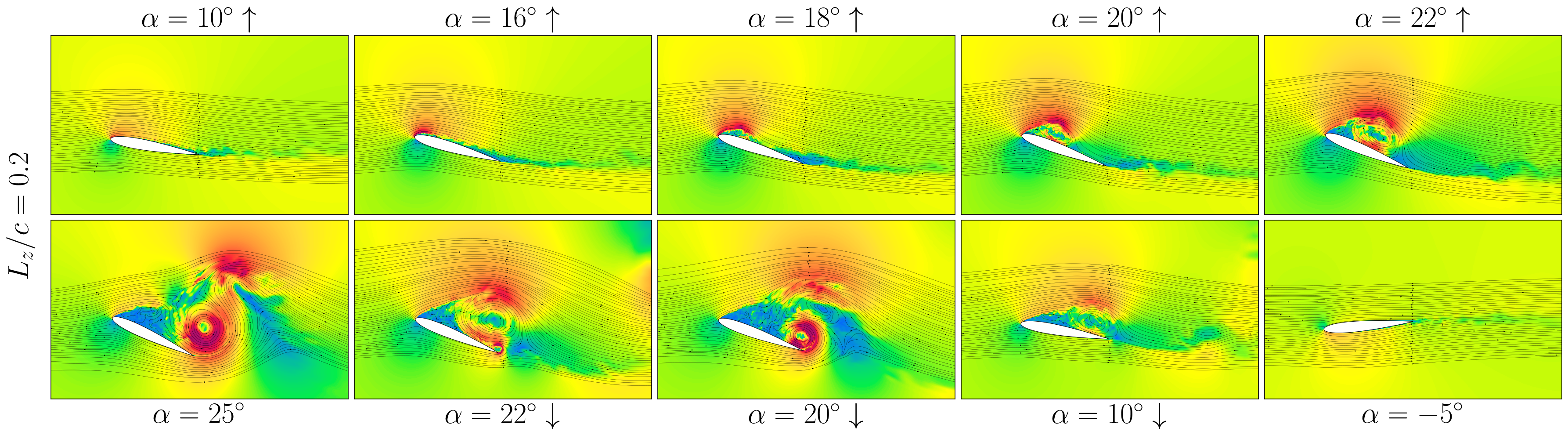}\\
	\end{subfigure}
	
	\vspace*{4mm}
	\begin{subfigure}{\linewidth}
		\includegraphics[width=.95\linewidth]{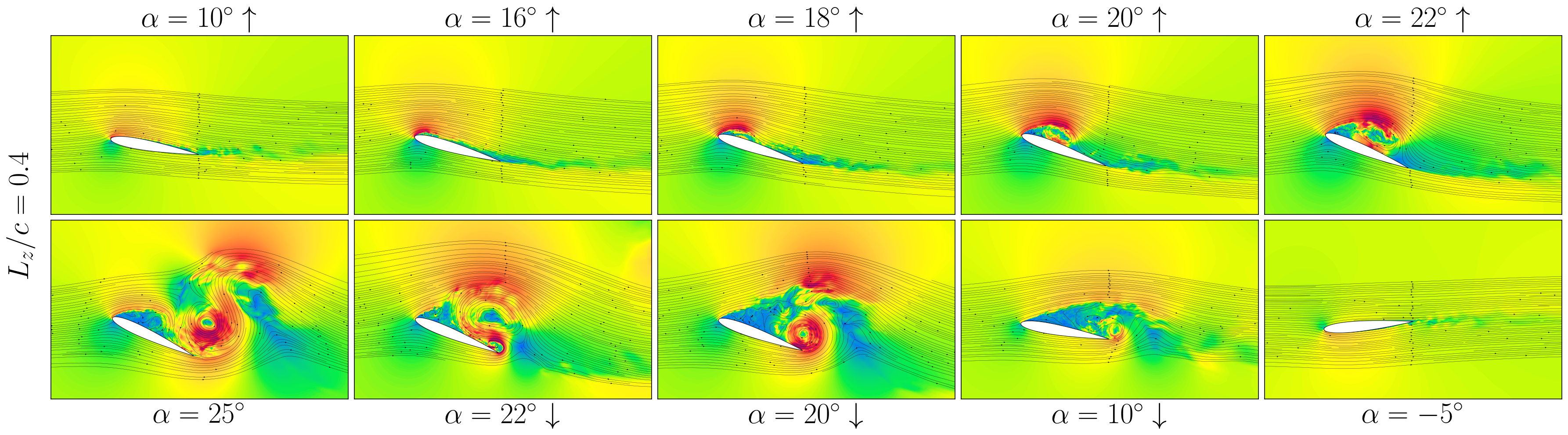}\\
	\end{subfigure}
	
	\vspace*{4mm}
	\begin{subfigure}{\linewidth}
		\includegraphics[width=.95\linewidth]{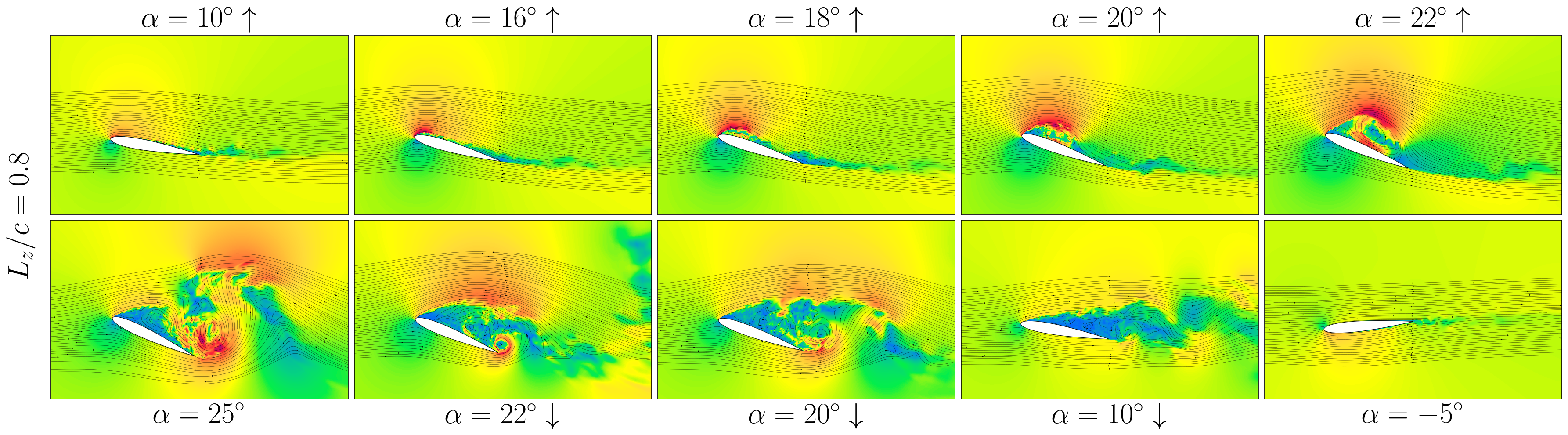}\\
	\end{subfigure}
	
	\vspace*{4mm}
	\begin{subfigure}{\linewidth}
		\includegraphics[width=.95\linewidth]{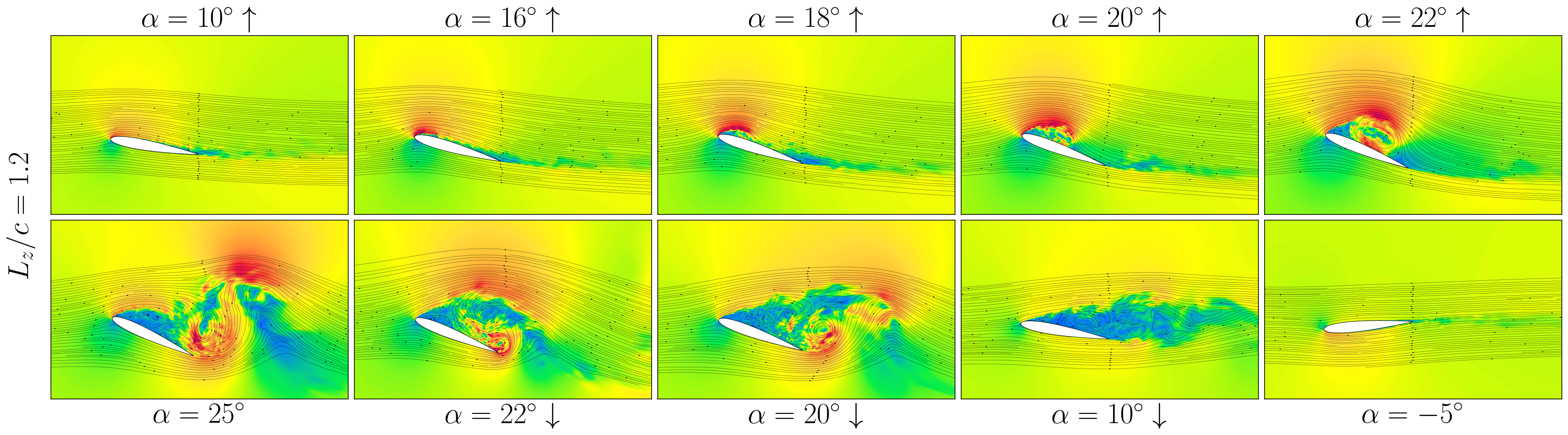}\\
	\end{subfigure}
	
	\vspace*{4mm}
	\subfloat{\includegraphics[width=0.4\linewidth]{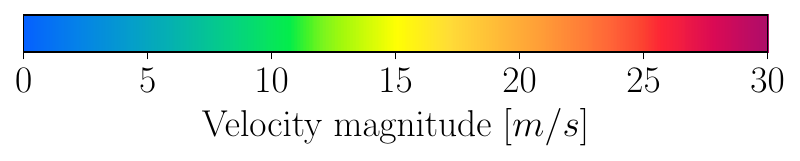}}
	
	\caption{Effect of span-to-chord ratio on instantaneous velocity magnitude contour at mid-section.}
	\label{img_model_comp_velocity_contour_les}
\end{figure*}

\begin{figure*}
	\centering
	\subfloat{\includegraphics[width=0.95\textwidth]{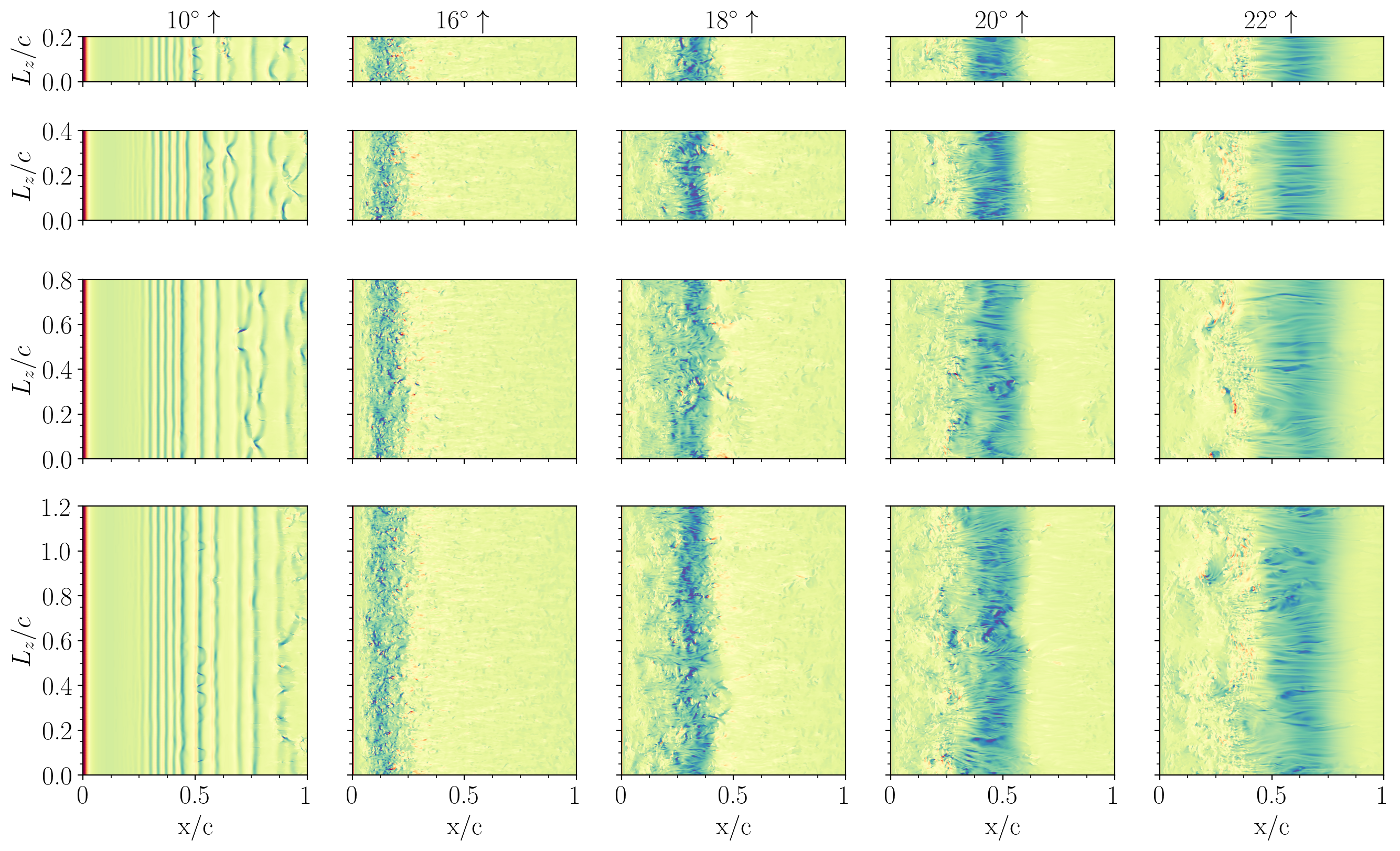}}\\
	\vspace*{2mm}
	\subfloat{\includegraphics[width=0.95\textwidth]{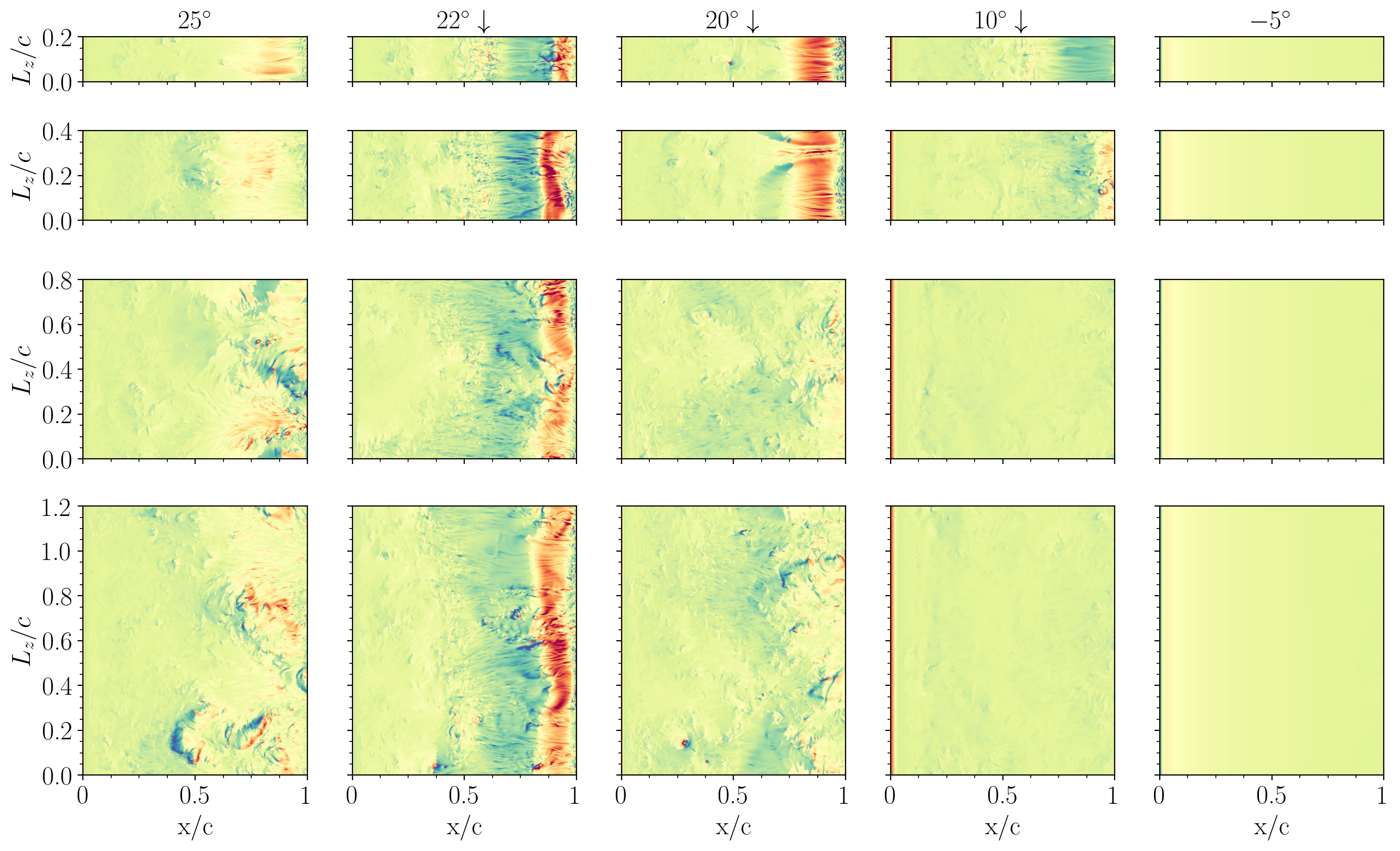}}
	\vspace*{2mm}
	\subfloat{\includegraphics[width=0.4\textwidth]{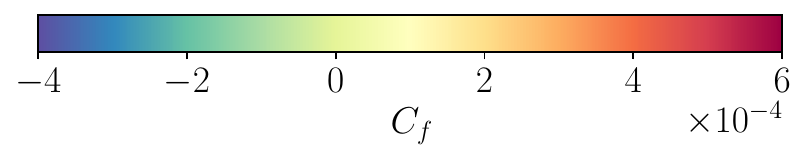}}
	
	\caption{Effect of span-to-chord ratio on instantaneous friction coefficient contour over the suction side of the airfoil.}
	\label{img_cf_les_comp}
\end{figure*}

\begin{figure*}
	\centering
	\subfloat[$L_z/c = 0.2$]{\includegraphics[width=0.24\textwidth]{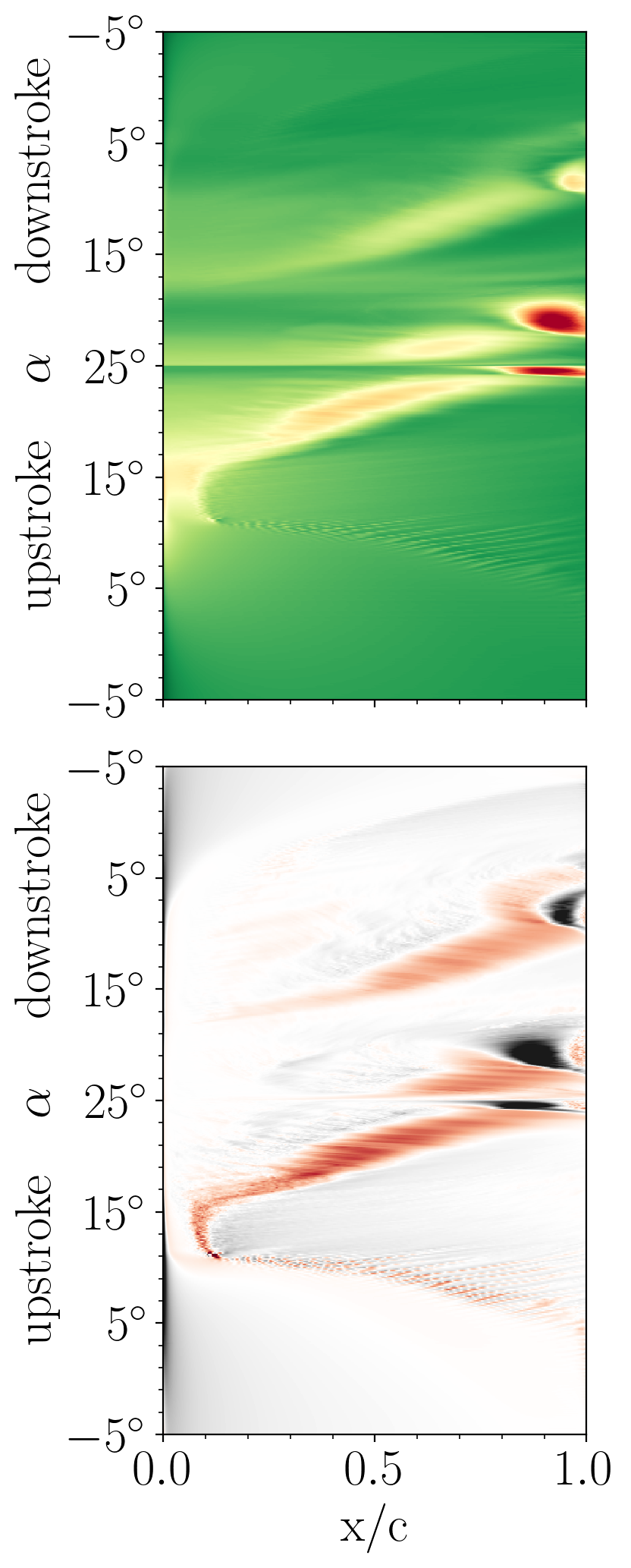}}
	\subfloat[$L_z/c = 0.4$]{\includegraphics[width=0.24\textwidth]{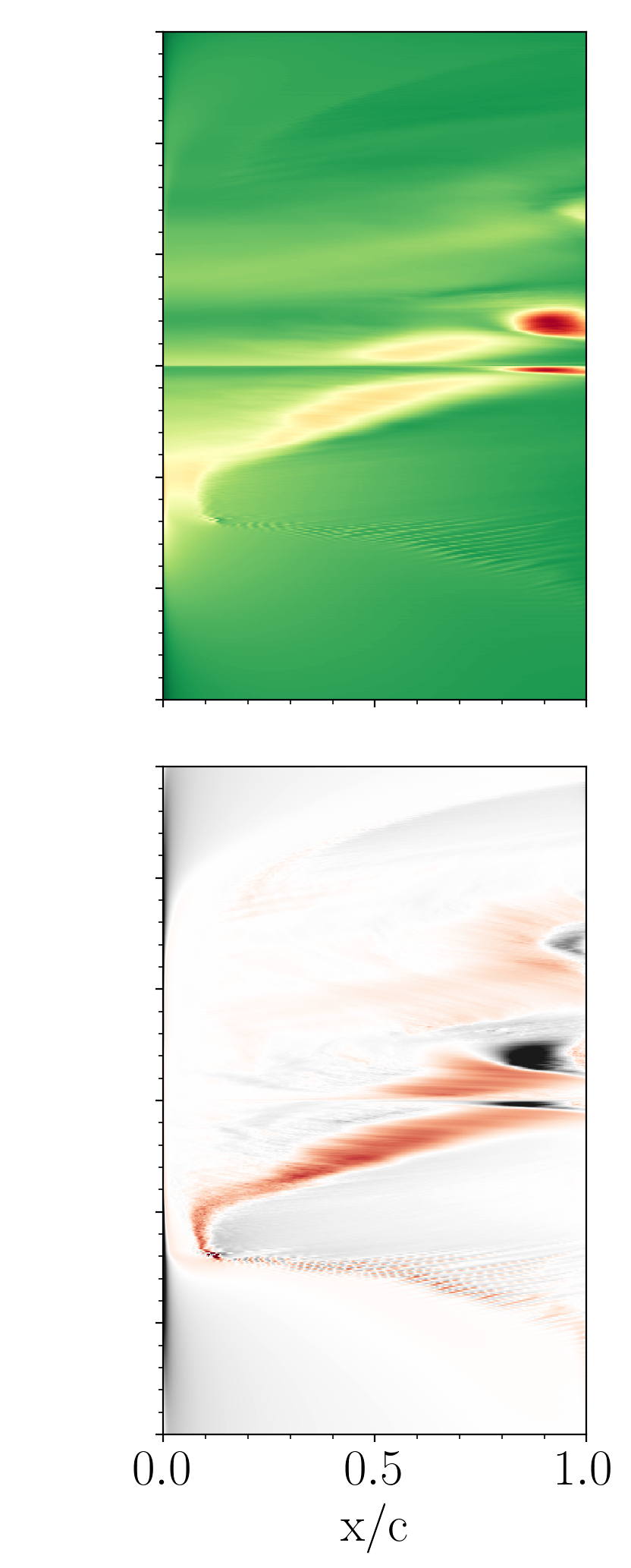}}
	\subfloat[$L_z/c = 0.8$]{\includegraphics[width=0.24\textwidth]{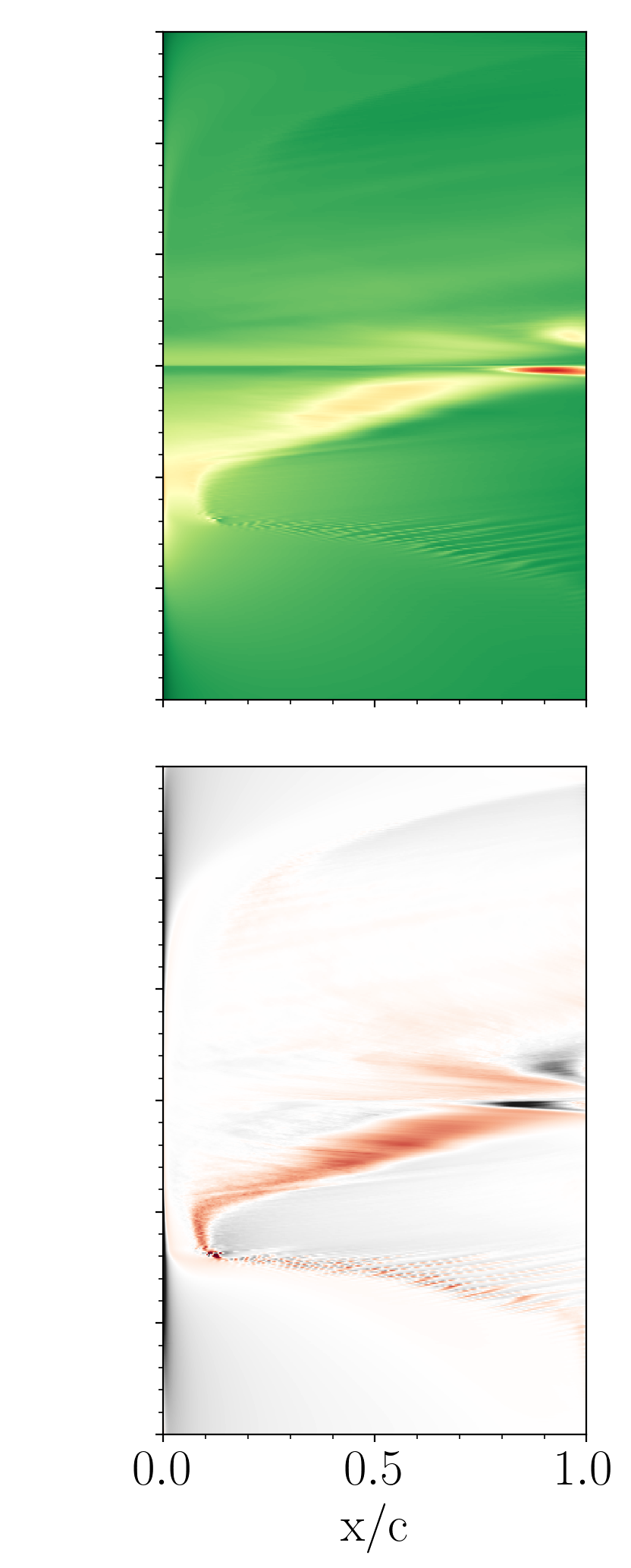}}
	\subfloat[$L_z/c = 1.2$]{\includegraphics[width=0.24\textwidth]{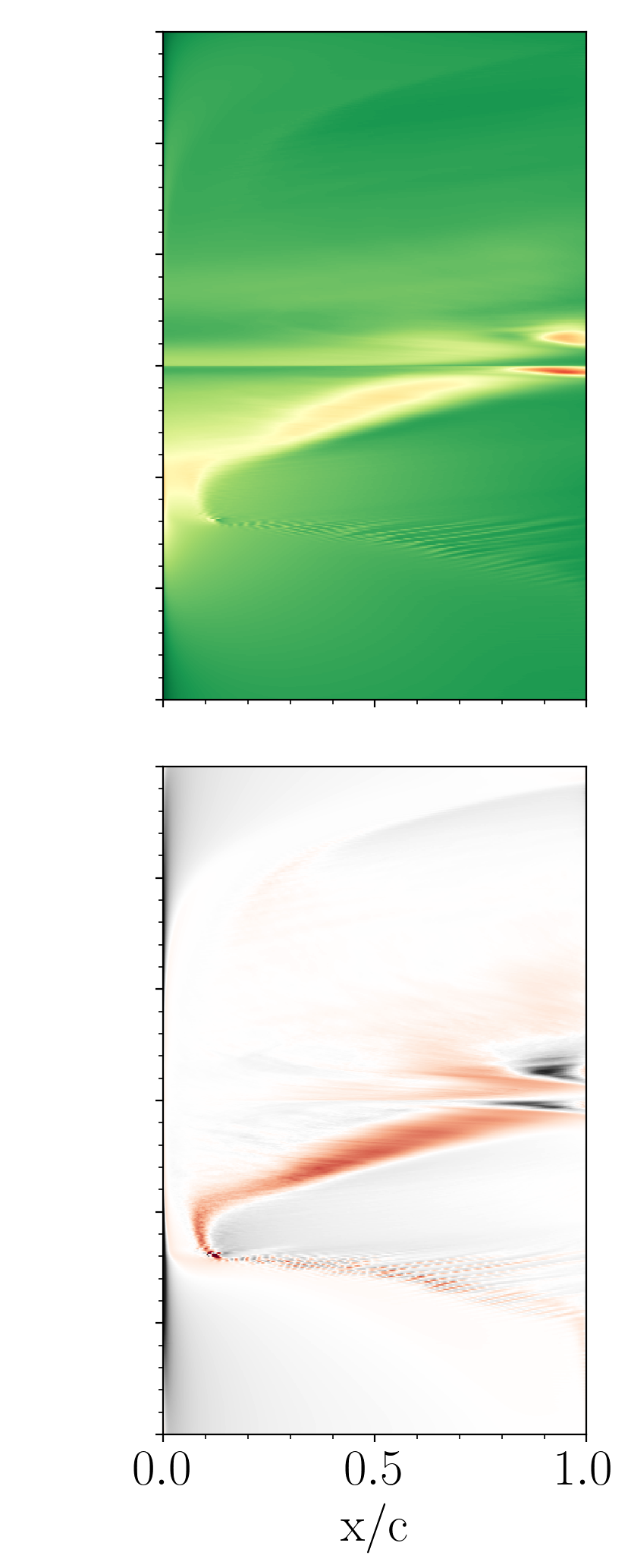}}\\
	\vspace*{5mm}
	\subfloat{\includegraphics[width=0.4\textwidth]{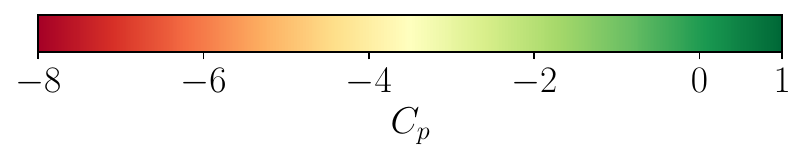}}
	\hspace*{1cm}
	\subfloat{\includegraphics[width=0.4\textwidth]{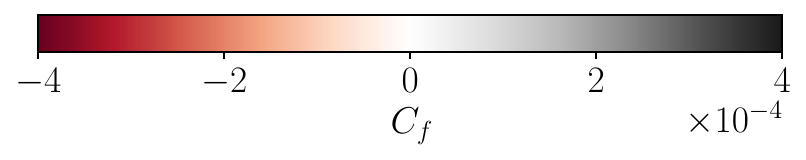}}
	
	\caption{Spanwise averaged pressure and friction coefficient over the suction side of the wing section.}
	\label{img_les_imshow}
\end{figure*}

\begin{figure*}
	\centering
	\includegraphics[width=\textwidth]{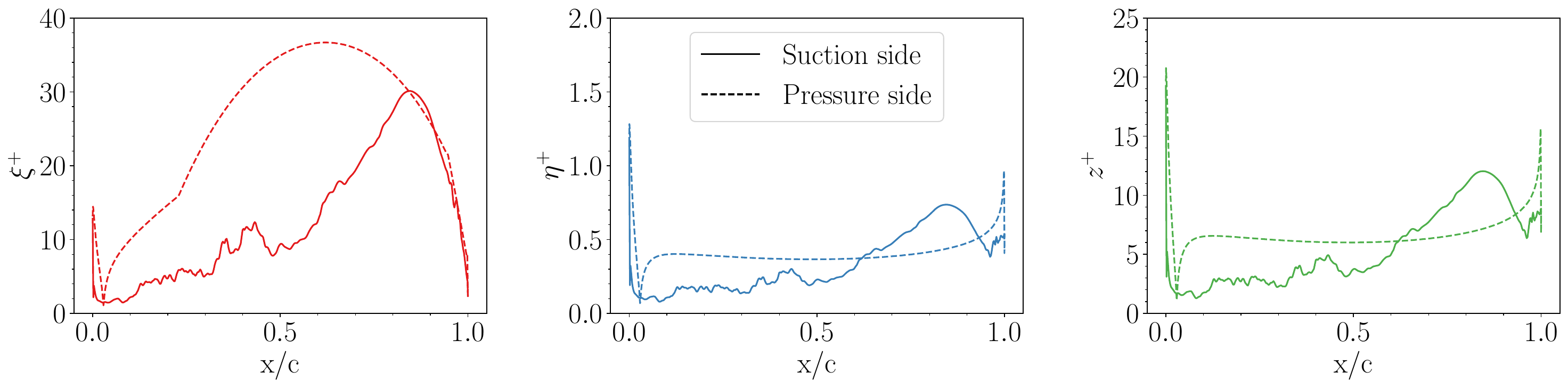}
	
	\caption{Spanwise averaged wall units at maximum incidence, $\alpha=25^\circ$.}
	\label{img_les_plus_unit}
\end{figure*}

\begin{figure*}
	\centering
	\subfloat[$\alpha=-0.6^\circ \uparrow$]{\includegraphics[width=0.45\textwidth,trim={0 3cm 0 0},clip]{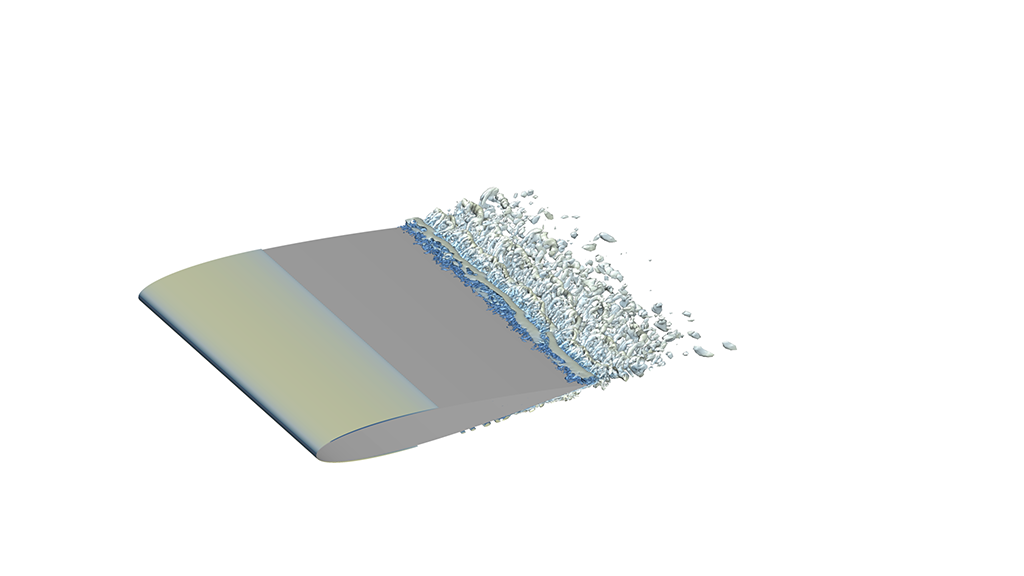}}
	\hfill
	\subfloat[$\alpha=10.0^\circ \uparrow$]{\includegraphics[width=0.45\textwidth,trim={0 3cm 0 0},clip]{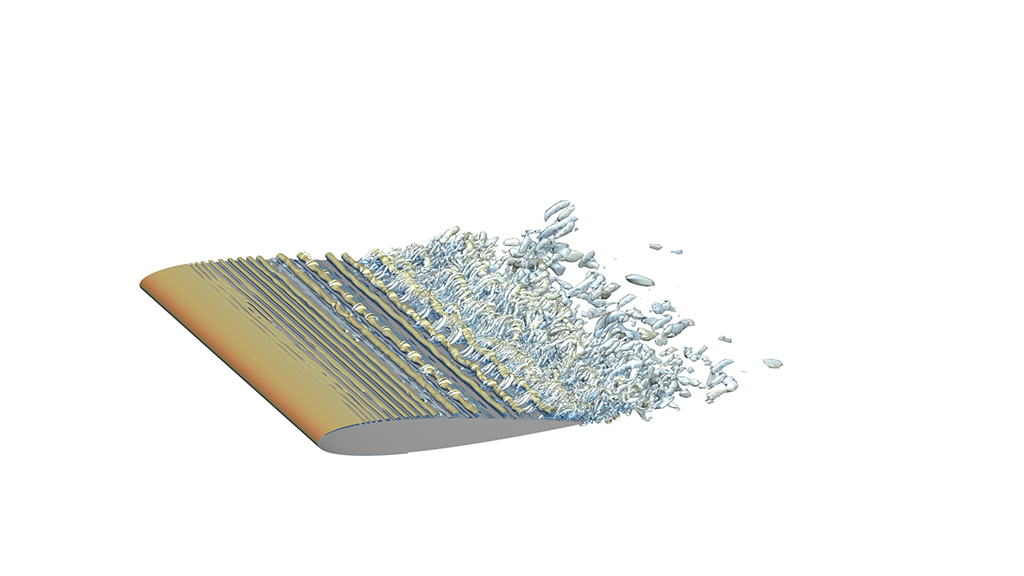}}
	
	\subfloat[$\alpha=20.6^\circ \uparrow$]{\includegraphics[width=0.45\textwidth,trim={0 3cm 0 0},clip]{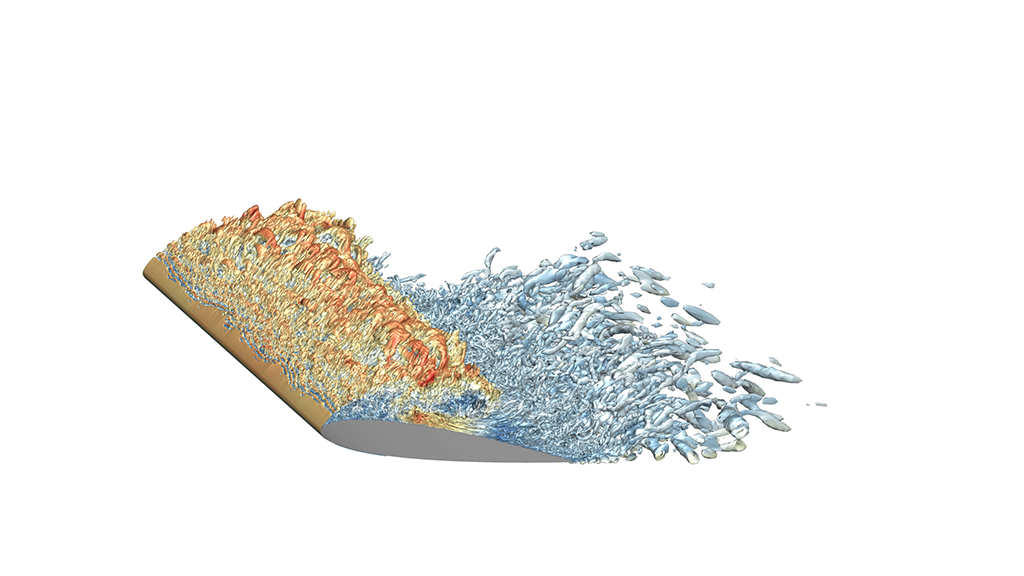}}
	\hfill
	\subfloat[$\alpha=25.0^\circ$]{\includegraphics[width=0.45\textwidth,trim={0 3cm 0 0},clip]{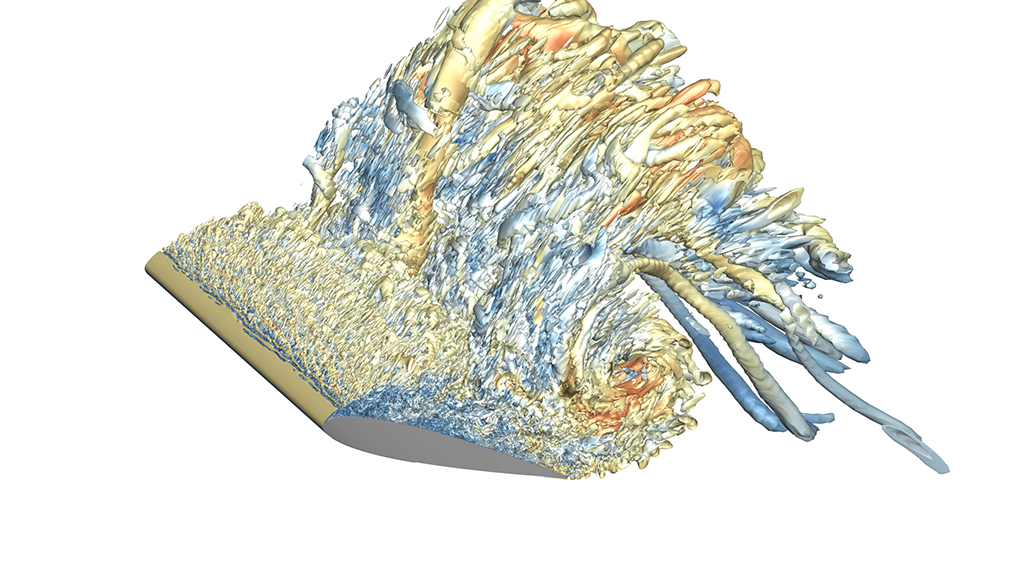}}
	
	\subfloat[$\alpha=20.6^\circ \downarrow$]{\includegraphics[width=0.45\textwidth,trim={0 3cm 0 0},clip]{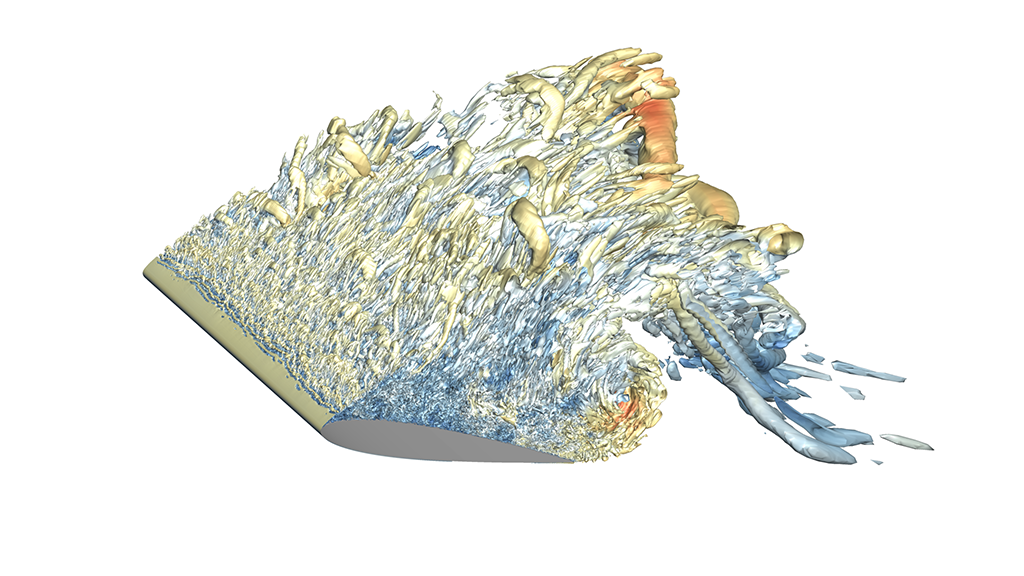}}\hfill
	\subfloat[$\alpha=10.0^\circ \downarrow$]{\includegraphics[width=0.45\textwidth,trim={0 3cm 0 0},clip]{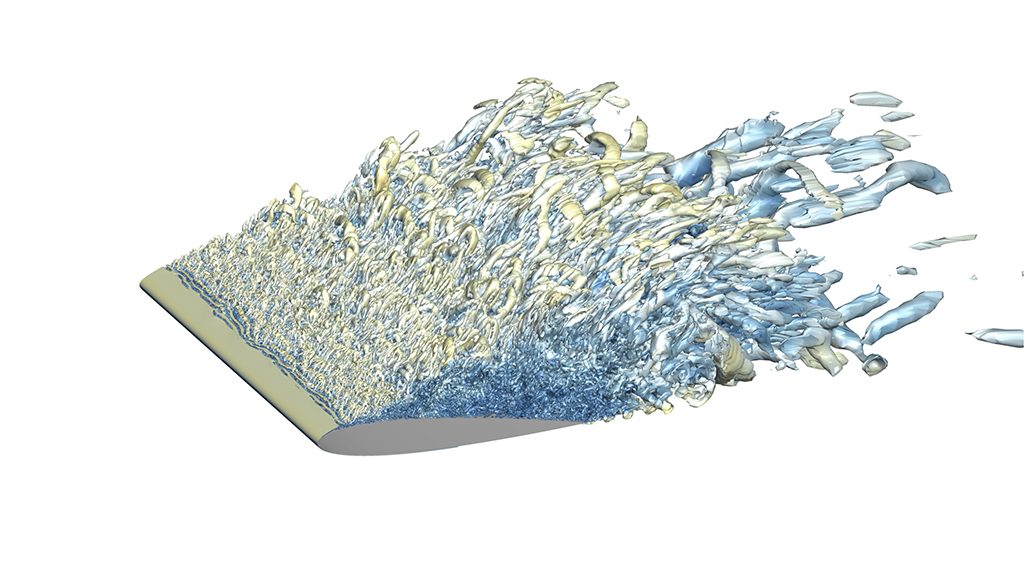}}
	
	\subfloat[$\alpha=-0.6^\circ \downarrow$]{\includegraphics[width=0.45\textwidth,trim={0 3cm 0 0},clip]{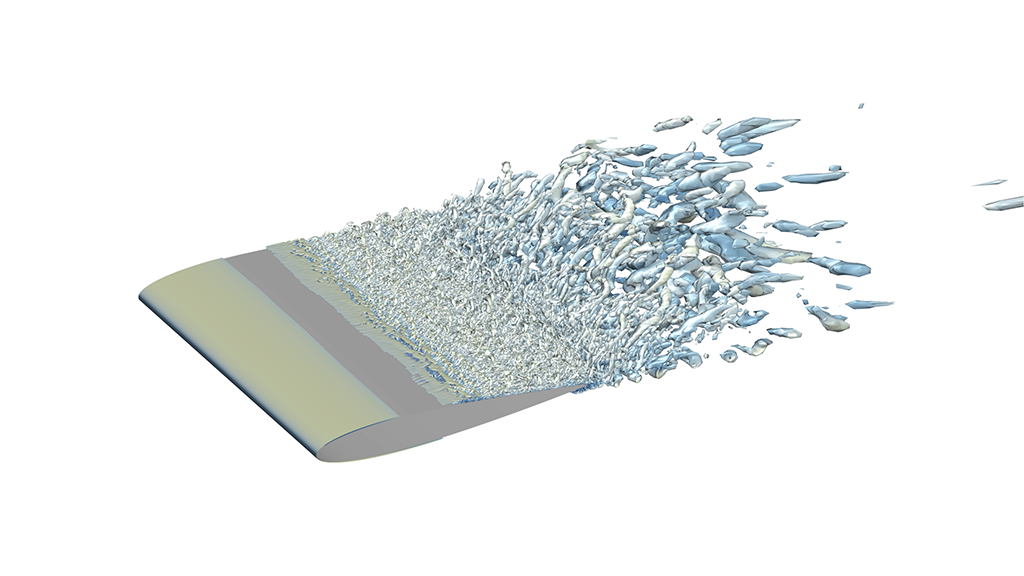}}
	\hfill
	\subfloat[$\alpha=-5.0^\circ$]{\includegraphics[width=0.45\textwidth,trim={0 3cm 0 0},clip]{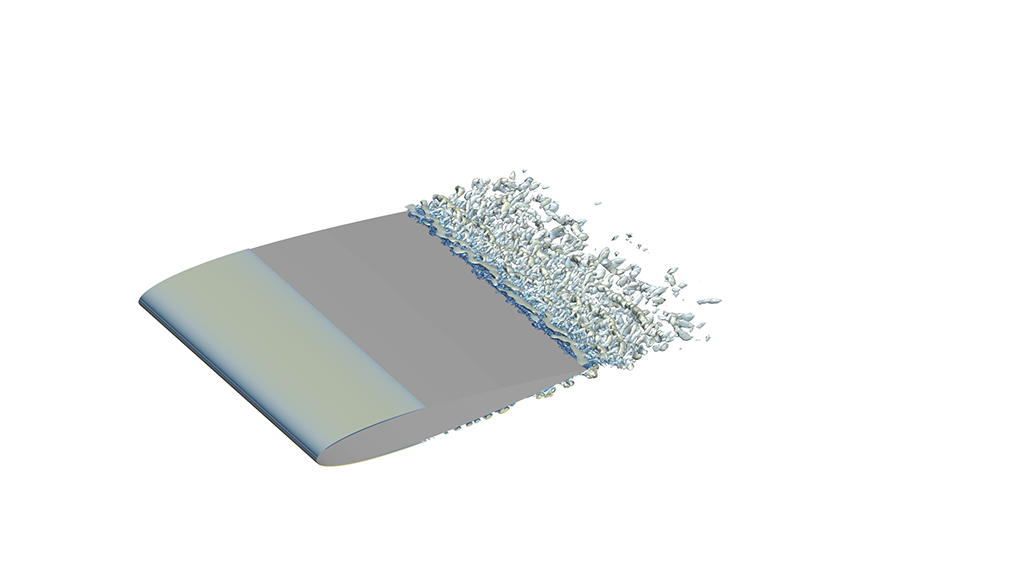}}
	
	\vspace*{5mm}
	
	\includegraphics[width=0.4\textwidth]{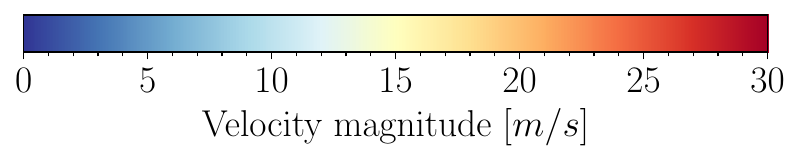}
	
	\caption{Q-criterion iso-surface with velocity magnitude contours at different instants of the pitching cycle. $\uparrow$ indicates the upstroke phase while $\downarrow$ the downstroke one. The model extends for $1.2c$ in spanwise direction.}
	\label{img_qcrit}
\end{figure*}

The initial stage of the investigation centers on the impact of employing wing geometries with limited span-wise dimensions to simulate dynamic stall phenomena.
Four different span-to-chord ratios are used: $L_z/c = 0.2, 0.4, 0.8, 1.2$.

The primary focus is on the aerodynamic loads experienced throughout the complete pitching cycle.
Figure~\ref{img_les_span} depicts the instantaneous lift and drag coefficients as functions of the angle of incidence.
If the flow remains predominantly attached, the behavior displayed at different span-to-chord ratios exhibits a high degree of similarity. 
During the initial portion of the upstroke, ranging from $\alpha = -5^\circ$ to $\alpha \approx 22^\circ$, the maximum difference of $C_L$ value is equal to 0.05.
However, a significant divergence is observed between the two smaller span-to-chord ratios when the LSB detaches and subsequently induces the DSV at an angle of incidence of approximately 22 degrees.
Notably, the first peak in lift during the pitch-down phase exceeds the values obtained with the larger span-to-chord ratios.
This disparity becomes even more pronounced at lower angles of attack, where the interaction between the TEV and the DSV results in a lift peak that even exceeds of 0.5 the $C_L$ value observed during the climb phase and of 0.3 the $C_D$ one. 
Lastly, larger grids exhibit a trend similar to the experimental data during pitch-down, although a $\approx 0.25 \ C_L$ offset is present.

Figure~\ref{img_model_comp_velocity_contour_les} illustrates the magnitude contours of instantaneous velocity at peculiar angles, providing a more detailed understanding of the influence of the span size.
The figure initially confirms that span-to-chord ratio has no impact on flow structures during the upstroke phase.
This is evident from the qualitatively similar appearance of the subfigures in the first row for each span-to-chord ratios.
Focusing on the angle of maximum incidence where the DSV is fully developed, we observe a significant effect of the span-to-chord ratio on the flow field.
For geometries with a ratio of 0.2 and even 0.4, the boundary conditions enforce a predominantly two-dimensional character on the flow.
This is manifested by a more defined DSV core and a higher peak velocity magnitude.
In contrast, geometries with larger spanwise extensions exhibit smaller turbulence scale structures, likely due to interactions in the spanwise direction.
This phenomenon becomes particularly pronounced during the descent phase of the airfoil.
Here, the geometries with smaller span-to-chord ratio exhibit a distinct vortical structure absent in the other cases.
A peculiar difference is seen at $\alpha$ equal to 10 degrees, where the two geometries with larger spans exhibit different behaviors.
The geometry with the largest span-to-chord ratio displays a completely stalled flow, whereas the geometry with a span of $0.8c$ still presents oscillatory patterns in the wake region.

Further insight into the span-to-chord ratio effect is provided by analyzing the skin friction coefficient distribution on the wing section's upper surface. Figure~\ref{img_cf_les_comp} presents the $C_f$ distribution for the previously investigated angles of attack across all four meshes.
The observations regarding the upstroke phase are corroborated by the alignment of the recirculation zone, identified by negative $C_f$ values.
The most significant discrepancies arise at the maximum angle of incidence, along with the descending phase at 20 and 10 degrees.
These contours demonstrate the development of three-dimensional structures within the flow field when sufficient spanwise extension is available.
Interestingly, a departure from prior observations is seen at 22 degrees, where the $C_f$ trends exhibit a high degree of similarity between all geometries. 
Additionally, the qualitative behavior at 10 degrees appears comparable for the two geometries with the larger span-to-chord ratios.

Figure~\ref{img_les_imshow} depicts the time-dependent evolution of the DSV and TEV by presenting the spanwise-averaged $C_p$ and $C_f$ trends on the suction side of the airfoil.
The boundary layer undergoing the transition from laminar to turbulent regime.
The transition starts at the trailing edge around an angle of attack of 2 degrees and progresses towards the LSB at 10 degrees.
Beyond 15 degrees, the development of the DSV becomes evident.
The DSV grows until it reaches the trailing edge at 22 degrees, where it interacts with the TEV.
The subsequent behavior varies depending on the grid size.
For geometries with $L_z/c$ of 0.2 and 0.4, two different pressure coefficient peaks are observed at the trailing edge, followed by a weaker third one.
Additionally, the detachment of another vortex from the mid-chord region is visible.
In contrast, the behavior for the geometries with larger spans is different.
The trailing edge experiences a pressure coefficient peak, albeit lower in magnitude compared to the smaller span-to-chord ratios, upon the first interaction with the DSV.
This is followed by a weaker peak, and subsequently, the flow becomes completely stalled, exhibiting no discernible trends.

To assess the resolution of the computational mesh, Figure~\ref{img_les_plus_unit} presents the wall units averaged in the spanwise direction for the computational mesh of the geometry with the largest span extent, evaluated at the maximum angle of incidence, 25 degrees.
The reported values adhere to the recommended practices for WRLES.
However, the $\eta^+$ value is lower than that imposed by LES best practice and is closer to 1. 
This value satisfies the requirement imposed by the RANS turbulence model adopted in the flow initialization.

Figure~\ref{img_qcrit} presents a visualization of the three-dimensional dynamic stall structures using the Q-criterion iso-surface with velocity magnitude contours.
A key observation is the difference in the fields at identical angles of attack during the ascent and descent phases.
Analyzing each phase, the laminar-to-turbulent boundary layer transition is visible during the 10-degree in the upstroke phase, characterized by the presence of instabilities and hairpin structures.
At an angle of approximately 20 degrees, the DSV begins to develop, exhibiting a primarily two-dimensional structure.
As the angle of attack reaches its maximum, where the DSV is fully developed and interacts with the TEV, the flow is completely three-dimensional and exhibits spanwise structures.
The descent phase at 20 and 10 degrees is still fully stalled and three-dimensional.
Note that this portion of the pitching cycle is not simulated in previous studies of dynamic stall that use a ramp-up motion, but is critical for describing dynamic stall in its entirely.
Furthermore, it is the condition when the largest differences occur depending on the model span size.
At the end of the descent phase, as the angle of attack approaches zero, the flow starts to laminarize.
Finally, at the lowest incidence, the flow presents some instabilities on the lower side of the wing section.

\subsection{Comparison with RANS and hybrid RANS/LES approaches}\label{ssec_other_simulations}

\begin{figure*}
	\centering
	\subfloat{\includegraphics[width=0.48\textwidth]{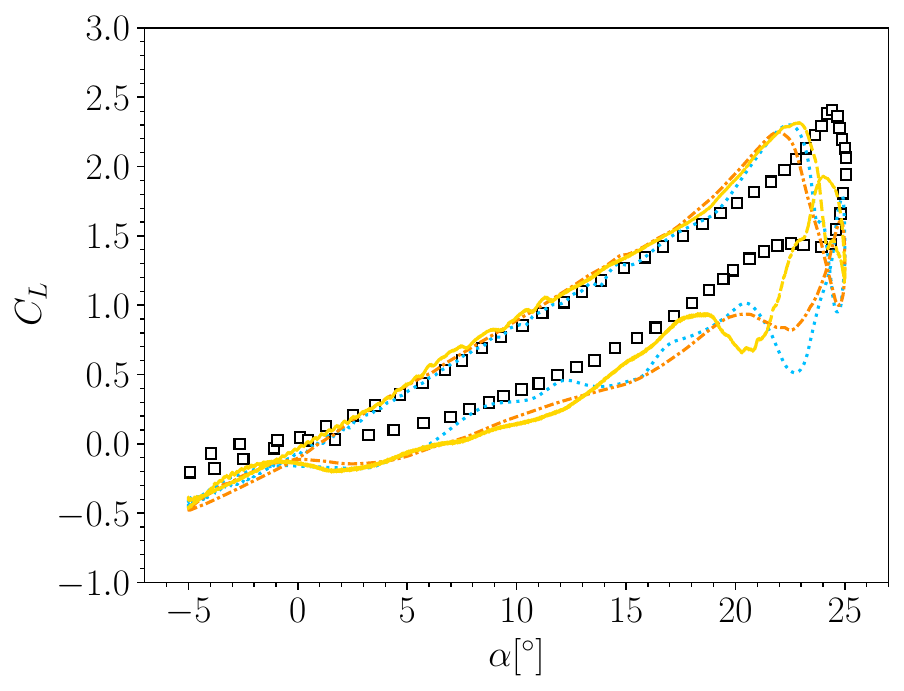}}
	\hfill
	\subfloat{\includegraphics[width=0.48\textwidth]{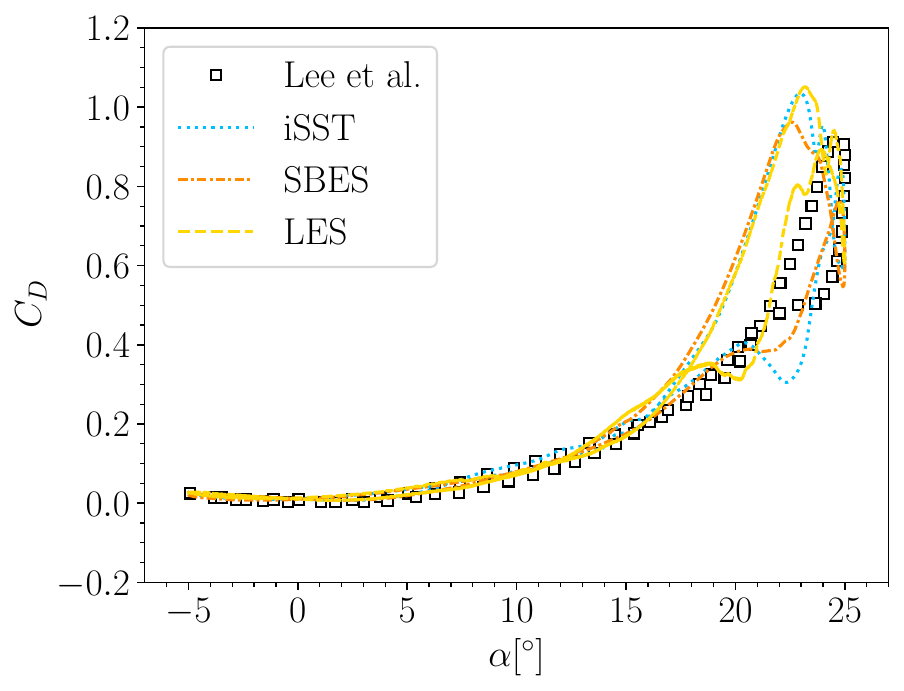}}\\
	
	\caption{Numerical simulation comparison of lift and drag coefficients. LES refers to the $1.2c$ spanwise case. SBES and iSST data from~\citet{Baldan2024b}. Experimental data from~\citet{Lee2004}.}
	\label{img_sim_comp}
\end{figure*}

We present now a comparative study of the LES results from the previous section with those obtained from previous studies by~\citet{Baldan2024b}.
Two approaches were employed by ~\citet{Baldan2024b}: a two-dimensional unsteady RANS simulation and a three-dimensional approach utilizing hybrid RANS/LES techniques.
For the RANS simulation, the $k-\omega$ SST turbulence model was implemented in conjunction with the $\gamma$ intermittency model, referred to as iSST, to enhance the description of the laminar-to-turbulent transition within the boundary layer.
The hybrid model employed a Stress-Blended Eddy Simulation approach coupled with the same RANS turbulence model.
The reference data for the LES simulations is based on the mesh with a span-to-chord ratio of 1.2.

Figure~\ref{img_sim_comp} presents a comparison of the profile loads obtained from the numerical simulations and the experimental data by~\citet{Lee2004}.
The $C_L$ results present a maximum 0.1 difference between the numerical models for attached flow conditions, with their curves nearly overlapping.
Notably, the most significant discrepancy between the numerical and experimental data lies in the angle of attack corresponding to peak lift which is $22^\circ$ and $24^\circ$, respectively.
Interestingly, the simulations anticipate, with respect to the experimental data, the angle at which the DSV is convected downwards and this behavior is consistent across all numerical models, including the WRLES approach. 
In contrast, during the initial descent phase of the profile $\alpha = 24^\circ - 18^\circ$, characterized by fully stalled flow, the models exhibit some divergence.
However, as the angle of incidence decreases, they converge towards the same solution.
Note that in the stall portion cycle-to-cycle variations are not considered due to the large computational resource requirements of large eddy simulations.
Despite a $C_L$ offset equal to $0.2$ from the experimental data persists, the overall behavior of simulations aligns well with the experiments.

\section{Conclusions}\label{sec_conclusion}

This work investigated the impact of the span-to-chord ratio on pitching airfoils using wall-resolved large eddy simulations.
The study revealed a critical dependency of key flow structures, particularly the three-dimensionality of the dynamic stall vortex, on the spanwise size.
While existing literature reports case studies with limited span-to-chord ratios, from 0.1 up to 0.2, these were found not adequate to capture the full complexity of the phenomenon.
Consequently, results obtained with such geometries are can be accepted only in the attached flow region, typical of ramp-up motions, involving the laminar separation bubble formation and the dynamic stall vortex generation. On the contrary, a complete pitching motion was used in the present work to highlight these limitations.
These results point to the growing challenge of studying the phenomenon with high-fidelity models at increasing Reynolds numbers due to the ever-increasing mesh size requirements.

Furthermore, a comparison with computational models based on RANS equations suggested a surprising result. 
The key discrepancies between numerical simulations and experimental data are not primarily due to turbulence models, as commonly reported. 
These discrepancies seem instead to be linked to other factors, such as possibly wind tunnel effects or the inherent limitations in accurately replicating the surface roughness of the wind tunnel wing section model within the computational models.

\section*{Acknowledgments}
	The authors acknowledge Leonardo SpA -- Helicopter Division for granting access to the \textit{davinci-1} supercomputer.

\end{document}